\documentstyle[aps,prl,twocolumn,psfig]{revtex}

\draft
\begin{document}
\newcommand {\be}{\begin{equation}}
\newcommand {\ee}{\end{equation}}
\newcommand {\bea}{\begin{eqnarray}}
\newcommand {\eea}{\end{eqnarray}}
\newcommand {\nn}{\nonumber}

\twocolumn[\hsize\textwidth\columnwidth\hsize\csname@twocolumnfalse%
\endcsname

\title{Phase Diagram of the quadrumerized Shastry-Sutherland Model} 

\author{Andreas L\"{a}uchli, Stefan Wessel, and Manfred Sigrist}   
\address{Institut f\"ur Theoretische Physik,
ETH-H\"onggerberg, CH-8093 Z\"urich, Switzerland
 }

\date{\today}
\maketitle

\begin{abstract}

We determine the phase diagram of a generalized Shastry-Sutherland model,
using a combination of  dimer- and quadrumer-boson methods and numerical exact
diagonalization techniques. Along special lines in the parameter space the model reduces
to  the standard Shastry-Sutherland model, the 1/5-th depleted square lattice
and the two-dimensional plaquette square lattice model. We study the evolution of the ordered
phases found in the latter two unfrustrated models under the effect of
frustration. Furthermore we present new exact diagonalization results for
the Shastry-Sutherland model on clusters with up to 32 sites, 
supporting the existence of an intermediate gapped valence bond crystal 
phase with plaquette long-ranged order. 
\end{abstract}

\vspace{1cm}
]

There has been considerable interest in recent years in the study of
low-dimensional quantum spin systems, both experimentally and theoretically. 
Special attention has been devoted to two-dimensional antiferromagnetic
systems, where quantum fluctuations and frustration allow for various exotic
quantum phases to compete with quasi-classical long-range magnetic order.\cite{review}

Already in the unfrustrated regime such spin liquid phases occur in certain
regions of the parameter space, a prominent example being the spin-1/2 Heisenberg model
on the 
1/5-th depleted square lattice.\cite{lee} In this model two distinct spin liquid phases,
well described by resonating valence bond (RVB) -like states, are found along with an
intermediate long-ranged antiferromagnetically ordered phase.\cite{lee,katoh}  The location of the quantum
critical points separating these phases are known to rather high precision.\cite{troyer1} 

When considering frustrated systems, ground state properties are
less well established. A prominent example of recent interest, due to its
relevance for the spin gap system SrCu$_2$(BO$_3$)$_2$,\cite{kageyama,miyahara} is the
Shastry-Sutherland model (SSM).\cite{shastry}  In this model the nearest-neighbor square
lattice antiferromagnet is frustrated by additional diagonal interactions,
arranged in
a staggered pattern on alternate squares. This model retains long-range
N\'{e}el-order for small diagonal coupling. Furthermore, it becomes an exact
dimer valence bond solid (VBS) with  singlets forming on the diagonal bonds for small 
axial coupling. Concerning the existence and nature of intermediate
phases, despite numerous investigations, 
[8-12], %\cite{albrecht, koga, chung,carpenties, weihong} 
a definite picture has not yet
emerged. 

\section{The model}
In this paper, we present exact diagonalization studies on the SSM that 
indicate the occurrence of a valence bond
crystal (VBC) in the intermediate regime, with plaquette long-range order. Furthermore, we
are able to link this phase to a consistent phase
diagram of an extension of the SSM. Therefore, we introduce the quadrumerized SSM,
defined on the square lattice by the following spin-1/2 Heisenberg
Hamiltonian
\begin{equation}
H = 
 K\!\! \sum_{{\langle i,j \rangle}_{\Box}} {\bf S}_i \cdot
{\bf S }_j+J \!\!\sum_{{\langle i,j \rangle}_{\Box'}}  {\bf S}_i \cdot
{\bf S }_j + J'\!\! \sum_{\langle \langle i,j \rangle\rangle } {\bf S}_i \cdot
{\bf S }_j. 
\end{equation}
Here $K$ and $J$ are the two inequivalent nearest neighbour exchange couplings, whereas
$J'$ denotes the next nearest neighbour alternating dimer coupling. The various couplings
are displayed in Fig. 1. 
\begin{figure}[h]
\centerline{\psfig{figure=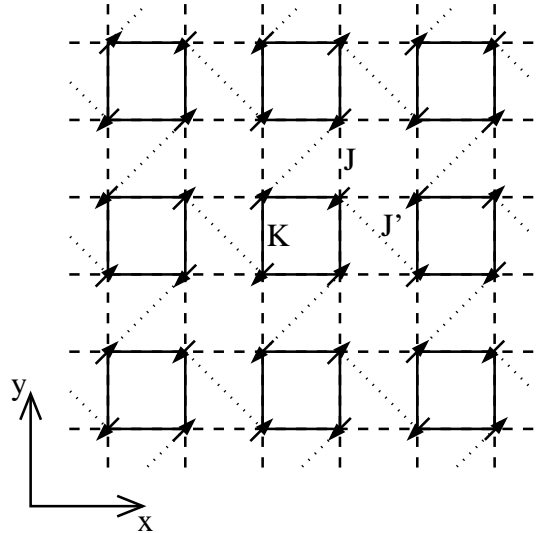,width=7cm,angle=0}}
\vspace{0.1cm}
\caption{
The quadrumerized Shastry-Sutherland lattice. A spin-1/2 degree of freedom is
located on each vertex. The various couplings, represented by different line
styles, are denoted $K$ (solid), $J$ (dashed), and $J'$ (dotted). The lattice is self-dual
under the exchange ($J \leftrightarrow K$). Arrows illustrate the ordered
phase found for $J=0$ in the region $J'\approx K$.  
}
\end{figure}
Note, that the lattice is self-dual under the
exchange ($J \leftrightarrow K$), hence only the case $J \leq K$ will be
considered.
Furthermore, the standard Shastry-Sutherland model is recovered along $J=K$, and  
has a larger space group symmetry. Other lines of enhanced
symmetry correspond to the 1/5-th depleted square lattice ($J=0$), and the
plaquette square lattice ($J'=0$), respectively.

The paper is organized as follows:
In the next section we present the  phase diagram of the quadrumerized SSM as obtained from
boson operator mean-field theory. Then in the third section we use exact
numerical diagonalization to study the
effects of frustration in the model and relate the numerical results to the
mean-field phase
diagram. We concentrate on the standard SSM in the
fourth chapter and provide evidence for a VBC intermediate
phase. A summary and conclusions are given in the final section.
   
\section{Boson operator approach}

We first review the numerical results obtained along the unfrustrated lines in the
parameter space of the Hamiltonian (1).
For the 1/5-th depleted square lattice ($J=0$) there
exists a plaquette RVB-like phase (PRVB) at small $J'/K <
0.94$, and a dimer RVB-like phase (DRVB) at large
$J'/K>1.67$, with an intermediate long-range ordered antiferromagnetic
phase.\cite{troyer1} The classical configuration corresponding to this order is
depicted by the arrows in Fig. 1. 

The plaquette square lattice ($J'=0$) is unfrustrated as well. Using
stochastic series expansion quantum Monte Carlo simulations a
quantum critical point is found at $(J/K)_c \approx 0.55$, separating a PRVB
spin liquid for $J/K < (J/K)_c$ from a gapless N\'{e}el-ordered phase.\cite{wessel} This value agrees well with results from
perturbation expansions.\cite{plaquettelattice} 

While quantum Monte Carlo proves powerful for studying the
unfrustrated limits of the Hamiltonian (1), due to the sign problem
other methods are needed,  once frustration is present.  
In order to study possible instabilities of the spin liquid phases, we use
standard 
boson operator mean-field theory, which is known to work on a qualitative
level even for large
frustration. Our analytical results will also be substantiated by
the numerical approach of the following sections. 

\subsection{Dimer-boson approach}

 Consider first the DRVB regime
$J' \gg J,K$ where the dimer-boson technique can be applied. \cite{sachdev}  In this
representation, the spin-1/2 degrees of freedom on each $J'$-dimer are
expressed by bosonic bond operators,
\begin{eqnarray}
\nonumber |s\rangle&=& s^{\dagger}|0\rangle =|0,0\rangle,\\
\nonumber |t_+\rangle&=& t_+^{\dagger}|0\rangle =|1,-1\rangle,\\
|t_0\rangle&=& t_0^{\dagger}|0\rangle =|1,0\rangle,\\
\nonumber |t_-\rangle&=& t_-^{\dagger}|0\rangle =|1,+1\rangle,
\end{eqnarray}
where the $|S,S^z\rangle$ denote the states on a given dimer. From the action
of the spin operators ${\bf S}_i$, $i=1,2$ (denoting the two sites of a
dimer) on these states the representation of the spin operators can be deduced
\begin{eqnarray}\label{spinoperatorsdimer}
\nonumber S^z_i&=&\frac{1}{2}\left( t^{\dagger}_+ t_+ -  t^{\dagger}_- t_-\right)
  -\frac{(-1)^i}{2} \left(t^{\dagger}_0 s + s^{\dagger} t_0\right),\\
S^{\pm}_i&=&\frac{1}{\sqrt{2}}\left( t^{\dagger}_{\pm} t_0 +  t^{\dagger}_0
  t_{\mp}\right)\pm \frac{(-1)^i}{\sqrt{2}}\left( t^{\dagger}_{\pm} s -
  s^{\dagger} t_{\mp}\right).
\end{eqnarray}
The spin commutation relations for ${\bf S}_i$, $i=1,2$ are obtained when the
bond operators obey bosonic statistics. Furthermore, the number of physical
states available on each dimer specifies a hard-core constraint for the
bosons on each dimer $\mu$,
\begin{equation}\label{hardcore}
s_\mu^{\dagger}s_\mu+t_{\mu ,+}^{\dagger}t_{\mu ,+} + t_{\mu ,0}^{\dagger}t_{\mu ,0} +
t_{\mu ,-}^{\dagger}t_{\mu ,-} = 1.
\end{equation}
Taking the interdimer couplings $J$ and $K$ into account and using Eq. (\ref{spinoperatorsdimer}) , the Hamiltonian (1)
is mapped onto an equivalent bosonic Hamiltonian,
$H^D=H^D_0+H^D_{I}$, containing quadratic, diagonal terms
\begin{eqnarray}\nonumber
H^D_0=J' \sum_{\mu} -\frac{3}{4} s^{\dagger}_{\mu} s_{\mu} + \frac{1}{4} \left( t^{\dagger}_{\mu+} t_{\mu,+} + t^{\dagger}_{\mu,0} t_{\mu,0}
  + t^{\dagger}_{\mu,-} t_{\mu,-}\right),
\end{eqnarray}
and quartic terms, $H^D_I$, describing the interdimer scattering. The square
lattice of dimers, with two dimers per unit cell, is found to reduce to a
square lattice with a single site per unit cell in the bosonic
representation. Here, we 
first imagine rotating half of the $J'$-dimers clockwise, so that all dimers align
along the  (1,-1)-direction (cf. Fig. 1). Then we
take the centers of
the dimers as the sites and use a coordinate system where $x_D$ is along the
original (1,-1)-direction and $y_D$ along (1,1).
To proceed, we need to
implement the constraint (\ref{hardcore}) by means of a Holstein-Primakoff
representation, \cite{chubukov}
\begin{eqnarray}\nonumber
s^{\dagger}=s=\sqrt{1-t^{\dagger}_+ t_+ - t^{\dagger}_0 t_0
  - t^{\dagger}_- t_-}, 
\end{eqnarray}
and then decouple the quartic $H^D_I$
  via a linear
approximation, similar to linear spin wave theory. The resulting total
quadratic Hamiltonian, ${{\bar H}}^D$, is then diagonalized in momentum space using a generalized Bogoliubov
transformation.\cite{bogo} This approach is expected to work well inside the DRVB phase.
A threefold degenerate spectrum of triplet excitations is
obtained, consistent with unbroken SU(2) symmetry,
\begin{equation}
\omega({\bf k}_D)=\sqrt{J'\left[J'+(K-J)(\cos{k_x}-\cos{k_y})\right]}.
\end{equation} 
Here, the wave vector ${\bf k}_D=(k_x,k_y)_D$ is defined with respect to the dimer
coordinate system $(x_D,y_D)$.
The phase boundaries of the
DRVB phase are obtained from the instabilities of the triplet excitation spectrum,
i.e. by a vanishing spin gap at ${\bf k}_D=(\pi,0)_D$,
signaling the condensation of the corresponding bosons at this wave vector.\cite{sachdev}
Mapping back onto the original square lattice, the corresponding magnetic order
is obtained, characterizing the phase beyond the instability line $2(K-J) \geq
J'$. In fact,
this magnetic order corresponds to the long-range order found for the 1/5-th depleted square
lattice, c.f. Fig. 1.    

\subsection{Quadrumer-boson approach}

When the parameters in the Hamiltonian of Eq. (1) are close to another
limiting case, $J,J' \ll K$, a similar approach, the
quadrumer-boson technique, can be applied.\cite{starykh}
 The Hamiltonian of a single quadrumer, $H=K [ {\bf S}_1 \cdot
{\bf S }_2 + {\bf S}_2 \cdot
{\bf S }_3 + {\bf S}_3 \cdot
{\bf S }_4 + {\bf S}_4 \cdot
{\bf S }_1 ]$, can be expressed in terms of the total spin, ${\bf S}={\bf
  S}_1+{\bf S}_2+{\bf S}_3+{\bf S}_4$, and the total subspin on each diagonal,  
 ${\bf S}_A={\bf
  S}_1+{\bf S}_3$, and  ${\bf S}_B={\bf S}_2+{\bf S}_4$. The spectrum is given
by $E(|S,S^z,S_A,S_B
\rangle)=K/2 (S^2-S_A^2-S_B^2)$. The lowest lying triplet
$\{|1,S^z,1,1\rangle, S^z=0,\pm1\}$ has
a gap, $\Delta=K$, to the ground state $|0,0,1,1\rangle$. Since there is
a further  gap, $\Delta'=K$, to the higher excitations, we attempt to obtain
the instabilities of the PRVB phase by using a restricted quadrumer-boson method,
omitting all the higher excitations on the quadrumers. Hence, the spin-1/2 degrees of
freedom on each quadrumer are expressed by bosonic operators on the
restricted Hilbert space,
\begin{eqnarray}
\nonumber |s\rangle&=& s^{\dagger}|0\rangle =|0,0,1,1\rangle,\\
\nonumber |t_+\rangle&=& t_+^{\dagger}|0\rangle =|1,-1,1,1\rangle,\\
|t_0\rangle&=& t_0^{\dagger}|0\rangle =|1,0,1,1\rangle,\\
\nonumber |t_-\rangle&=& t_-^{\dagger}|0\rangle =|1,+1,1,1\rangle.
\end{eqnarray}
From the action of the spin operators ${\bf S}_i$, $i=1,...,4$ in the
restricted Hilbert space, the following representation can be deduced\cite{footnote1}
\begin{eqnarray}\label{spinoperatorsquadrumer}
\nonumber S^z_i&=&\frac{1}{4}\left( t^{\dagger}_+ t_+ -  t^{\dagger}_- t_-\right)
  -\frac{(-1)^i}{\sqrt{6}} \left(t^{\dagger}_0 s + s^{\dagger} t_0\right),\\
S^{\pm}_i&=&\frac{1}{2\sqrt{2}}\left( t^{\dagger}_{\pm} t_0 +  t^{\dagger}_0
  t_{\mp}\right)\pm \frac{(-1)^i}{\sqrt{3}}\left( t^{\dagger}_{\pm} s -
  s^{\dagger} t_{\mp}\right).
\end{eqnarray}
In the restricted Hilbert space, the hard-core constraint (\ref{hardcore}) is now
obeyed on each quadrumer. 
Expressing the Hamiltonian (1) in terms of the quadrumer-boson operators, a
bosonic Hamiltonian, $H^P= H^P_0 + H^P_I$, is obtained with a noninteracting
diagonal part
\begin{eqnarray}\nonumber
H^P_0=K \sum_{\mu} -2\: s^{\dagger}_{\mu} s_{\mu} - \left( t^{\dagger}_{\mu+} t_{\mu,+} + t^{\dagger}_{\mu,0} t_{\mu,0}
  + t^{\dagger}_{\mu,-} t_{\mu,-}\right),
\end{eqnarray}
and a quartic scattering part $H^P_I$. Here the sum extends over the square lattice of
quadrumers
formed by the $K$-bonds in Fig. 1. Following the decoupling procedure already
used in the dimer-boson approach, the following threefold degenerate triplet excitation
spectrum is obtained in the PRVB regime,
\begin{equation}
\omega({\bf k}_P)=\sqrt{K\left[K-\frac{2}{3}(2J - J')\left(\cos{k_x}+\cos{k_y}\right)\right]}.
\end{equation} 
The minimum of this spectrum is located at ${\bf k}_P=(\pi,\pi)_P$ when $J' >2 J$,
and the gap vanishes for $J'>3/4(K+2J)$, corresponding again to the order depicted
in Fig. 1. Furthermore, for $J' < 2 J$ the minimum is located at ${\bf
  k}'_P=(0,0)_P$, and the gap is again closed for $J' < 2J - 3/4 K$. In this regime
the model becomes long-range N\'{e}el-ordered.
 
Upon comparing the ground state energies from the dimer- and the quadrumer-boson
approach inside the common range of stability, we can obtain the direct
first order transition line between the DRVB and PRVB spin liquid phases.\cite{starykh}
\begin{figure}
\centerline{\psfig{figure=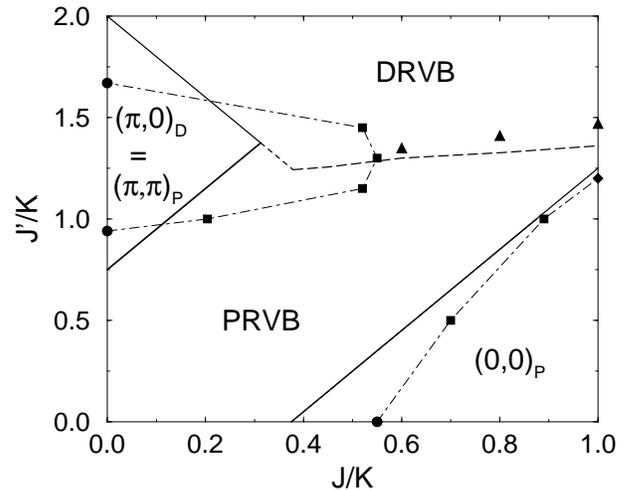,width=8cm,angle=270}}
\caption{
Phase Diagram of the quadrumerized Shastry- Sutherland model. Ordered phases are characterized by the ordering wave vectors in the
boson representations. Solid lines indicate second order transition lines, and the
dashed line the first order transition. Also shown are points on the phase
boundaries from quantum Monte
Carlo (circles), [4,13] exact
diagonalization (squares, and triangles), and series expansion
(diamond). [12] Dashed-dotted lines are guides to the eye. The increment of the
parameter scan in exact diagonalization was $\Delta(J/K)=0.1$ 
[$\Delta(J'/K)=0.2$] for the $(\pi,\pi)_P$ [$(0,0)_P$] phase boundary.} 
\end{figure}
The overall phase diagram is shown in Fig. 2.
The spin liquid phases
are characterized by the corresponding RVB-like state, while we label the long-range ordered
phases by the ordering wave vectors in the boson operator approaches. 
Furthermore, solid lines represent
second order phase transitions, whereas the
dashed line indicates the first order transition line.

The N\'{e}el-ordered phase $(0,0)_P$ extends up to rather large frustration,
with a largest extend of  $(J'/J)_{\mathrm{max}}\approx 1.2$ along the
Shastry-Sutherland line ($J=K$). On the other hand, the largest extent of the
$(\pi,\pi)_P$ phase, for $J'/K\approx 1.3$ is bound
by $(J/K)_{\mathrm{max}} < 0.55$ from exact diagonalization.
This difference can be  traced back to
the ratio of the number of frustrating couplings to the number the initial couplings, which is
$1:4$ when starting at $J'=0$, but $2:3$ upon starting at $J=0$.

Furthermore, from the phase diagram in Fig. 2, we find that
the DRVB phase of the 1/5-th depleted square lattice is
adiabatically connected to the exact dimer VBS phase of the standard
Shastry-Sutherland model (the dimer VBS state fails to be an exact
eigenstate for
$J\neq K$).
On the other hand, the DRVB phase is not adiabatically connected to the PRVB,
as expected on topological grounds.\cite{bonesteel} Hence, we  find a first order phase transition
separating the two spin liquid phases beyond the regime of the $(\pi,\pi)_P$
phase.

When turning to the case $J>K$, the phase diagram shown in Fig. 2 is 
obtained upon  
interchanging $J$ and $K$, due to the invariance of the Hamiltonian in Eq. (1)
under the exchange $(J\leftrightarrow
K)$. Furthermore, we label
the plaquette RVB-like phase for  $J>K$ by PRVB$'$, since now singlets are predominantly
formed on a different set of quadrumers than in the PRVB phase.

\section{Exact Diagonalization Studies}
 We include in Fig. 2
 the positions of quantum critical points along the unfrustrated lines,
 obtained by quantum Monte Carlo.\cite{troyer1,wessel} 
These compare rather well with
the above mean-field theory. 
To extend the numerical analysis into the frustrated regime we have performed
exact diagonalization studies on clusters with $N=8,16,32$ spins, using
periodic boundary conditions,  along various lines
in the phase diagram. We determine the finite size values of the order
parameter, $M'$, defined by \cite{poilblanc}
\begin{equation}\label{fsz1}
M'^2(N)=\frac{1}{N(N+2)}\left\langle \left( \sum_{i} \epsilon_i {\bf S}_i \right)^2 \right\rangle.
\end{equation}
Here $\epsilon_i$ takes on the values $\pm 1$ at site $i$, according to
the
pattern in Fig. 1 for the $(\pi,\pi)_P$ phase or the standard N\'{e}el-order for the
$(0,0)_P$ phase, respectively. 
Using the finite-size data we determine $M'$ from the scaling law \cite{poilblanc}
\begin{equation}\label{fsz2}
M'(N) = M' +  \frac{c_1}{N^{1/2}} + \frac{c_2}{N}.
\end{equation}
For example, in Fig. 3 we show  results obtained along the line $J'=1.3$, where
quantum Monte Carlo simulations at $J=0$ give a maximum moment of $M'\approx 0.23$.\cite{troyer1} Within exact
diagonalization we can reproduce this value, and furthermore observe a smooth
decrease in $M'$ upon increasing the frustration, up to a critical point at
$J/K \approx 0.55$, where we enter into the spin liquid regime. Proceeding in a
similar fashion we obtain the critical points depicted by squares in
Fig. 2. Moreover, along the line $J'=2J$ no finite order parameter was obtained
after finite size scaling. 
\begin{figure}[t]
\centerline{\psfig{figure=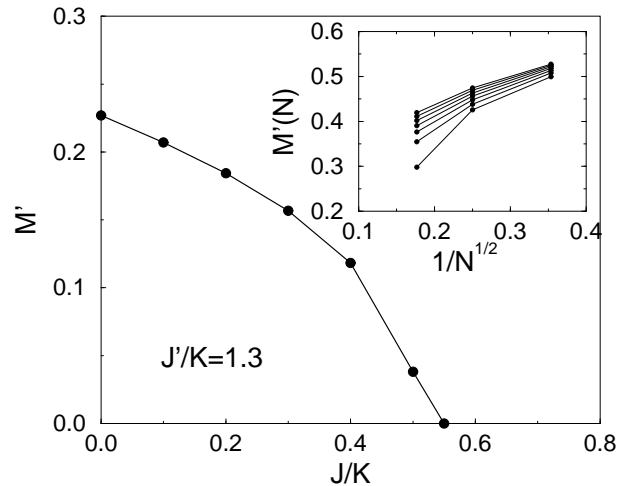,width=8cm,angle=270}}
\caption{
Evolution of the order parameter of the $(\pi,\pi)_P$ phase upon increasing
$J/K$ for $J'/K=1.3$. %obtained using Eqs. (\ref{fsz1},\ref{fsz2}).
The inset showns the finite size data obtained for
$N=8,16,32$ at $J/K$ $=$ $0,0.1,...,0.6$, (top to bottom). 
}
\end{figure}
 
The two  spin liquid phases cannot be separated within exact
diagonalization using space group symmetry. Namely, upon increasing $J'/K$
for a constant $J/K$, the representation class of the ground state does not
change. Nevertheless, from the approximate slopes of the ground state energy
vs. $J'/K$ at constant $ J/K$ in
the regions of small and large $J'/K$ respectively, we estimated the first order transition
points indicated by triangles in Fig 2. 

Comparing these numerical results with the mean-field calculations in the last
section, we conclude that the boson operator
approach gives a good qualitative account of the phase diagram of the quadrumerized
SSM. Namely, the characterization of the various phases and the
location of the phase transition lines agree well with numerical results.

\section{Shastry-Sutherland model}

In the quadrumer-boson approach we find a finite window on the
Shastry-Sutherland line ($J=K$), where both plaquette spin liquid phases, PRVB
and PRVB$'$, 
come arbitrary close
to the $J=K$ line. This already indicates an intermediate phase in the
standard SSM between the N\'{e}el-ordered phase and the
dimer VBS  phase. Similar conclusions were obtained in a field theoretical study of a
generalized SSM, which does not break the symmetry needed
by the dimer VBS state to be an exact
eigenstate.\cite{carpenties} We now focus on this intermediate regime in the
standard SSM.

There has recently been considerable interest in the nature of the
intermediate phase. In the large spin, classical limit
the system retains N\'{e}el-order for $J'/J\leq 1$ and is helically ordered otherwise,
with a twist between next nearest neighbour spins of $q =\arccos(-J/J')$.
Using Schwinger boson mean-field theory, Albrecht
and Mila predicted 
the existence
of a helical phase separating the dimer VBS and the ordered phase also for the 
spin-1/2 case, in a range $1.1
< J'/J < 1.65$.\cite{albrecht} 
Field-theoretical studies by Chung, Marston, and Sachdev for a generalized spin-S model with Sp(2N) symmetry suggest
the helical order to occur  in a larger range, $1.02 < J'/J < 2.7$, at
$S=1/2$.\cite{chung}  Furthermore, this approach predicts a phase with
plaquette order in the extreme quantum limit, $1/S > 5$. Indeed, using series expansions around the plaquette
limit of Eq. (1), Koga and Kawakami found the intermediate phase in a range
$1.16 < J'/J < 1.48$ to be
adiabatically connected to the PRVB phase.\cite{koga}  However, extended series expansions
by  Weihong, Oitmaa, and Hamer lead to different conclusions.\cite{weihong}  They suggest the
PRVB phase to become unstable before the Shastry-Sutherland line is reached and 
found a columnar dimer phase to be a possible candidate for the intermediate phase.
\begin{figure}[b]
\centerline{\psfig{figure=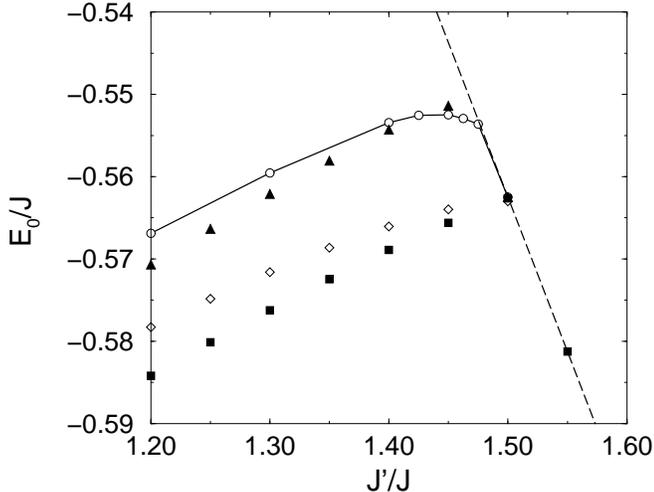,width=8cm,angle=270}}
\caption{
Ground state energy per site of the Shastry-Sutherland model from exact
diagonalization of clusters with $N=$ $16$ (squares), $20$ (triangles), $24$
(diamonds),  and $32$ (circles) sites. The solid line is a guide to the eye,
and the dashed line is the exact dimer VBS
state energy. 
}
\end{figure}
Here, we perform exact diagonalization studies on clusters with up to 32 sites,
significantly beyond the largest cluster sizes studied so far (24 sites in
Ref. \cite{totsuka}).
In
Fig. 4 we show the ground state energy of the SSM close to
the transition into the dimer VBS state for various cluster sizes. 
The solid line shows our results for the 32 site cluster, whereas other sizes are
represented by symbols. In addition, the energy of the 
dimer VBS state is shown by the dashed line.
Due to the different symmetry of the various clusters, the ground state enery
per site does not show a monotonous finite size dependence.
Nevertheless, independent of the
cluster size we find the system to be within the dimer VBS for
$J'/J >1.5$, consistent with the upper bound for the intermediate phase given
by Koga and Kawakami, but significantly below the values from the Schwinger
boson, and Sp(2N) theory.  From finite size analysis of the N\'{e}el-order parameter, we
estimate a upper bound for the ordered phase of
$(J'/J)_{\mathrm{max}}<1.4$, consistent with the series expansion value
$(J'/J)_{\mathrm{max}}=1.2 \pm 0.1$ \cite{weihong}. 
More interestingly, for the largest cluster we find a characteristic change in the
curvature of the ground state energy, well before entering the dimer VBS.
Hence, we conclude that characteristic features of the intermediate
phase could be retrieved from clusters with $N=32$ sites in a range $1.425 < J'/J
< 1.475$. 

In this regime we find the spin-spin, or 2-point correlation function to decrease
rapidly with distance, indicating the absence of antiferromagnetic order. 
\begin{table}
\begin{center}
\begin{tabular}{|c|c||c|c|}
  $(k,l)$ &  $C^4(0,16;k,l)$  & $(k,l)$ & $C^4(0,16;k,l)$ \\
\hline
25 , 26 &0.0213 &20 , 26& 0.0140 \\
29 , 30 &0.0222 & 21 , 27 & 0.0180 \\
31 , 9  & 0.0788 & 22 , 1  & 0.0274 \\
2 , 3   & 0.0825 & 17 , 23 & 0.0152 \\
6 , 7   & 0.0222 & 18 , 24 & 0.0148 \\
12 , 13 & 0.0209 & 8 , 14  & 0.0175 \\
8 , 28  & 0.0140 & 28 , 15 & 0.0239 \\
14 , 15 & 0.0144 & 29 , 31 & 0.0274 \\
21 , 22 & 0.0140 & 9 , 30  & 0.0152 \\
1 , 27  & 0.0153 & 4 , 10  & 0.0239 \\
4 , 5   & 0.0138 & 5 , 11  & 0.0175 \\
10 , 11 & 0.0160 & 6 , 12  & 0.0148 \\
17 , 18 & 0.0140 & 7 , 13  & 0.0180 \\
23 , 24 & 0.0214 & 19 , 25 & 0.0140 \\
19 , 20 & 0.0223 & 13 , 20  & -0.0124 \\
24 , 25 & -0.0107 & 14 , 21 & -0.0143 \\
28 , 29 & -0.0104  & 15, 22 & -0.0207 \\
15 , 31 & -0.0007 & 10 , 17 & -0.0019 \\
1 , 2   & -0.0007 & 11 , 18 & -0.0135 \\
5 , 6   & -0.0137 & 27 , 4  & -0.0177 \\
11 , 12 & -0.0107 & 1 , 5   & -0.0177 \\
30 , 4  & -0.0104 & 2 , 6   & -0.0019 \\
9 ,  10 & -0.0292 & 3 , 7   & -0.0207 \\
3 , 23  & -0.0292 & 23 , 8  & -0.0103 \\
7 , 8   & -0.0137 & 24 , 28 & -0.0164 \\
13 , 14 & -0.0163 & 25 , 29 & -0.0142 \\
20 , 21 & -0.0125 & 26 , 30 & -0.0135 \\
26 , 27 & -0.0163 & 12 , 19 & -0.0138 \\
18 , 19 & -0.0125 & & \\
\end{tabular}
\end{center}
\caption[99]{Dimer-dimer correlations $C^4(1,2;k,l)$  in the ground state of the
  Shastry-Sutherland model on the 32 sites cluster at $J'/J=1.45$. The
  labeling of the sites is shown in Fig. 5.
}
\label{table1}
\end{table}
In order to test against the various proposed ground states, we
measure the dimer-dimer, or 4-point correlation functions,
\begin{equation}
C^4(i,j,;k,l)=\langle {\bf S}_i \cdot {\bf S}_j \: \: {\bf S}_k \cdot {\bf
  S}_l \rangle - \langle {\bf S}_i \cdot  {\bf S}_j \rangle \langle {\bf S}_k \cdot
  {\bf S}_l \rangle,
\end{equation}
on the $N=32$ lattice in the above interaction range. In particular, we fix
$(i,j)=(0,16)$ and extend $(k,l)$ over all inequivalent $J$ bonds.
The values 
obtained in the ground state for $J'/J=1.45$  are displayed in Table I, and illustrated in Fig. 5, which
also shows the labeling of the sites. 
\begin{figure}[]
\centerline{\psfig{figure=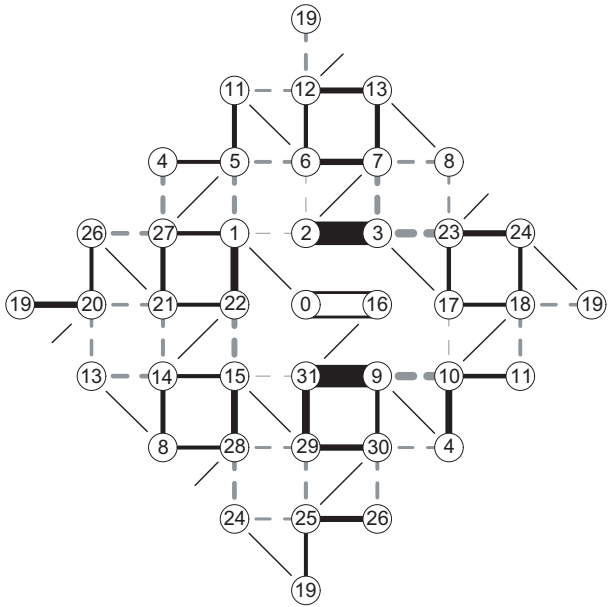,width=8cm,angle=0}}
\vspace{5mm}
\caption{
Dimer-dimer correlations in the ground state of the Shastry-Sutherland model
on the 32 sites cluster at $J'/J=1.45$. The reference bond is the bond
$(0,16)$. Positive (negative) correlations are drawn as full (dashed) lines. The
thickness of a line is proportional to the strength of the
correlation. Short diagonal lines indicate the position of the $J'$-dimer bonds. 
}
\end{figure}
We obtain a clear signal in the dimer-dimer correlations that  extends
throughout the whole cluster, with a finite asymptotic value approximately
reached for the larger dimer-dimer distances on the cluster. In the
spatial distribution we furthermore observe periodic oscillations, which reflect an
underlying order of quadrumer-singlets, formed predominately
on void squares (i.e. those squares not containing a diagonal bond).

For the SSM  there are two
equivalent configurations with quadrumer-singlet coverings residing on the two different
subsets of void squares (formed by the $J$ or $K$ bonds in Fig. 1,
respectively). 
Hence, if the plaquette-like order in the 4-spin correlation function survives
quantum fluctuations, indicative of a plaquette VBC, 
a two-fold degenerate ground state manifold will emerge in the thermodynamic limit. On a
finite lattice this degeneracy is lifted, but a low-lying
singlet state well inside the triplet gap and only slightly above the 
ground state energy is expected. 
We can obtain the quantum numbers of this
singlet state from the following symmetry considerations: The
Shastry-Sutherland lattice has a p4mm space group symmetry,\cite{knetter} and the ground
state of the 32 site
cluster has momentum ${\bf k}=(0,0)$, and is
invariant under the $\pi/2$ rotations about the center of any void square (i.e. s-wave like).
Furthermore, the two equivalent configurations of quadrumer-singlet coverings
are related by the reflections about either diagonal
dimer axes, but are invariant under the $\pi/2$ rotations and lattice
translations. Namely, they both reside already inside a single unit cell of
the Shastry-Sutherland lattice.

\begin{figure}[t]
\centerline{\psfig{figure=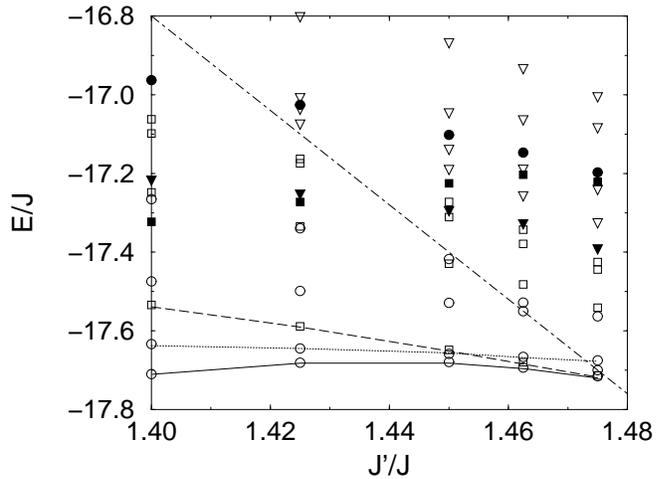,width=8cm,angle=270}}
\caption{
Ground state energy, and low-lying singlet and triplet excitations in the
${\bf k}=(0,0)$ sector of the
Shastry-Sutherland model on the 32 sites cluster. Open symbols represent
singlet states, full symbols triplet states. Circles denote states with eigenvalues
$R_{\pi/2}= 1$ (s-wave), squares denote $R_{\pi/2}=-1$ 
(d-wave), and  triangles $R_{\pi/2}=\pm i$ (two-fold degenerate). The solid line is a
guide to the eye for the ground state, the dotted to the lowest excited s-wave
state, and the dashed line the lowest d-wave state. The Dashed-dotted line is
the exact dimer VBS state  energy.
}
\end{figure}
Hence, a low-lying s-wave symmetric singlet state with momentum $(0,0)$ is
expected to be included 
in the spectrum of the 32 site cluster in the regime of the  plaquette VBS
phase. This state should furthermore show similar dimer-dimer correlations as 
the absolute ground state.

In Fig. 6 we plot the ground state energy along with those of the lowest excited singlet and triplet
states for the 32 site cluster in the zero momentum sector. We specify the
transformation properties of the various states under the $\pi/2$ rotation by the
eigenvalue $R_{\pi/2}= 1$ (s-wave), $-1$ (d-wave), or $\pm i$ (two-fold degenerate). See the caption of
Fig. 6 for a detailed account on the various used symbols. 
In the regime where we expect evidence for an
intermediate state we indeed find various low-lying singlet states well
inside the rather large triplet gap. Moreover, there are two singlet states with
energies rather close to the ground state energy, one being a s-wave,  and the
other a d-wave state with respect to the $\pi/2$ rotations. We
furthermore calculated the
dimer-dimer correlations for both states and find for the s-wave state a
similar signal as for the absolute ground state. 
On the other hand, the d-wave
state does not show any pattern in the dimer-dimer correlations. 
This state does not
seem relevant for the ground state of the system in the thermodynamic
limit. Presumably it is related to low-lying excitations in the dimer
VBS phase.\cite{totsuka,knetter}

From the above we conclude that the low-lying s-wave singlet state,
having the right quantum numbers and 
dimer-dimer correlation,  will become degenerate with the ground state upon
increasing the cluster size to the infinite lattice.
Both states then form the two-fold degenerate
ground state manifold of a plaquette VBC in the
thermodynamic limit. In Fig. 7 a pictorial description of this ground state
manifold is given. 
\begin{figure}[h]
\centerline{\psfig{figure=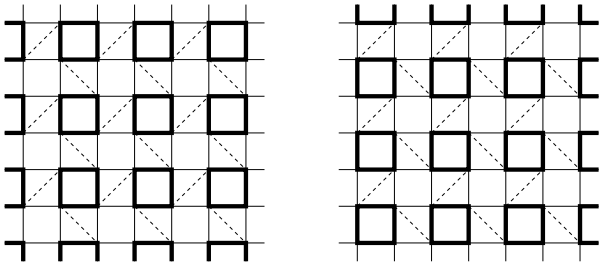,width=8cm,angle=0}}
\vspace{0.5cm}
\caption{
Pictorial illustration of the variational ground state manifold of the
Shastry-Sutherland model in the plaquette VBC phase. Thick lines
indicate four spins involved in a quadrumer-singlet. 
}
\end{figure}
Furthermore, this ground state manifold is invariant with
respect to lattice translations and the $\pi/2$ rotations about the centers of
the void squares, but each state spontaneously
breaks the reflection symmetry about the diagonal axes along the dimer directions.

From our numerical results we can also 
exclude a columnar dimer state, which would be four-fold degenerate and
furthermore show a different pattern in the dimer-dimer correlations than 
depicted in Fig. 5. \cite{leung}

Further evidence for the relevance of the plaquette VBC in the intermediate
regime of the SSM can be drawn from 
analogous results of
recent studies on the spin-1/2 Heisenberg model on 
the checkerboard lattice.\cite{fouet,honecker} In this model, diagonal bonds are again organized
on a square lattice in a pattern as to leave half of the
squares void. Namely, the underlying lattice is obtained by adding
in an additional diagonal bond on each square on which  a dimer bond is located
in Fig. 1.
Also in the checkerboard lattice quadrumer-singlets form on 
the void squares, resulting in a two-fold degenerate plaquette VBC ground state manifold
with broken space group symmetry. 
The structure leading to frustration is rather similar in both models, and
the system tries to minimize frustration by forming plaquette valence bonds on the
void squares. Due to the homogeneous axial couplings, this can be accomplished
only upon spontaneously breaking the symmetry inherited from the underlying lattice.

\section{Conclusions}
In conclusion, we have studied the phase diagram of the quadrumerized
Shastry-Sutherland model. Using bond-operator methods and exact numerical
diagonalization its phase diagram was established, which links the
various results available for special limiting cases of the
model. The antiferromagnetically ordered phase of the 1/5-th depleted square
lattice model is
destroyed by modest frustration, whereas the N\'{e}el-ordered phase
extends up to rather large frustration. There is a first order transition line
separating the different spin liquid phases, PRVB and DRVB, beyond the ordered
phase. Furthermore, the  DRVB is
adiabatically connected to the dimer VBS phase of the SSM.

For the standard
SSM there exists a finite region around $J'/J=1.45$
where the system becomes  
a plaquette VBC with spontaneously broken space group symmetry, and a two-fold
degenerate
ground state manifold. Perturbing away from the Shastry-Sutherland line,
i.e. for $J\neq K$, the
symmetry is broken explicitly, and the system favors a unique ground state,
namely the PRVB for $J<K$, and the PRVB$'$ for $J>K$. 
Furthermore, upon varying
$J'$ along the Shastry-Sutherland line, a
first order transition leads to the dimer VBS and a second order
transition to the N\'{e}el-ordered phase. Within our analytical and numerical
studies we did not find indications for further phases in the
(quadrumerized) SSM. 

Stabilization of a plaquette VBC phase in both the
Shastry-Sutherland and the checkerboard lattice model agrees with the generic
structure of the underlying frustrated lattice.
However, in the SSM the VBC is unstable towards the dimer VBS upon
increasing the diagonal  coupling.
For the checkerboard
lattice model the range of the VBC phase is still unkown, and remains for further studies.

We acknowledge fruitful discussions with Stephan Haas, Andreas Honecker, and
Bruce Normand. This work has been financially supported by the Swiss Nationalfonds.

{\it Note added.--} After completion of this work Akihisa Koga, and Norio Kawakami pointed us to
Ref. \cite{takushima}, where the phase diagram of the Hamiltonian in Eq. (1)
was studied using series expansion methods. Their results are in perfect
agreement with our calculations in sections II and III. However in \cite{takushima} the VBC nature of
the intermediate phase 
in the Shastry-Sutherland model was not noticed.

\end{document}